\documentclass[11pt,a4paper]{article}
\usepackage[utf8]{inputenc}
\usepackage{authblk}
\usepackage{amsmath}
\usepackage{amssymb}
\usepackage{mathtools}
\usepackage[pdftex]{graphicx}
\usepackage{float}
\usepackage{hyperref}
\usepackage[backend=biber,	
            natbib=true,
            style=numeric,	
            citestyle=numeric-comp, 
            sorting=none,
            giveninits=true,
            doi=false,
            hyperref=true
]{biblatex}

\usepackage{xcolor}
\usepackage{xspace}
\usepackage{hyphenat}
\usepackage[normalem]{ulem}

\numberwithin{equation}{section}

\newcommand{\alpS}{\ensuremath{\alpha_s}}
\newcommand{\alpSZ}{\ensuremath{\alpha_s(m_\PZ)}}
\newcommand{\aspi}{\ensuremath{a_s}}
\newcommand{\fbinv}{\ensuremath{\,\mathrm{fb}^{-1}}}
\newcommand{\GeV}{\ensuremath{\,\text{Ge\hspace{-.08em}V}}}
\newcommand{\muf}{\ensuremath{\mu_f}}
\newcommand{\mur}{\ensuremath{\mu_r}}
\newcommand{\MSbar}{\ensuremath{\overline{\textrm{MS}}}}
\newcommand{\pp}{\ensuremath{\mathrm{p}{}\mathrm{p}{}}}
\newcommand{\PQt}{\ensuremath{\mathrm{t}}}
\newcommand{\PAQt}{\ensuremath{\overline{\PQt}}}
\newcommand{\PZ}{\ensuremath{\mathrm{Z}\xspace}}
\newcommand{\mt}{\ensuremath{m_{\PQt}}}
\newcommand{\mtpole}{\ensuremath{m_{\PQt}^{\mathrm{pole}}}}
\newcommand{\mtMSbar}{\ensuremath{\overline{m}_{\PQt}}}
\newcommand{\mtMSR}{\ensuremath{m_{\PQt}^{\mathrm{MSR}}}}
\newcommand{\TeV}{\ensuremath{\,\text{Te\hspace{-.08em}V}}}
\newcommand{\ttbar}{\ensuremath{{\PQt{}\overline{\PQt}}}}

\newcommand{\Ndof}{\ensuremath{\mathrm{N}_\mathrm{dof}}}

\newcommand{\MCFM} {\textsc{MCFM}\xspace}

\title{
\vspace*{-30mm}
{\hfill \small{DESY 23-001}}\\[-1ex]
{\hfill \small{UWThPh-2023-1}}\\
\vspace{25mm}
Investigation of the scale dependence in the MSR and $\MSbar$ top quark mass schemes for the $\ttbar$ invariant mass differential cross section using LHC data}

\author[a,b]{Toni M{\"a}kel{\"a}\footnote{toni.makela@cern.ch}}
\author[c,d]{Andr{\'e} H. Hoang\footnote{andre.hoang@univie.ac.at}}
\author[a,e]{Katerina Lipka\footnote{katerina.lipka@desy.de}}
\author[f]{Sven-Olaf Moch\footnote{sven-olaf.moch@desy.de}}

\affil[a]{{\small Deutsches Elektronen-Synchrotron, Notkestr.~85, 22607 Hamburg, Germany}}

\affil[b]{{\small National Centre for Nuclear Research, Pasteura 7, PL-02-093 Warsaw, Poland}}

\affil[c]{{\small Faculty of Physics, University of Vienna, Boltzmanngasse 5, A-1090 Vienna, Austria}}

\affil[d]{{\small Erwin Schr\"odinger Institute for Mathematics and Physics, University of Vienna, Boltzmanngasse 9, A-1090 Vienna, Austria}}

\affil[e]{{\small Fakult{\"a}t f{\"u}r Mathematik und Naturwissenschaften, Bergische Universit{\"a}t Wuppertal, Gau{\ss}strassse 20, D-42119 Wuppertal, Germany}}

\affil[f]{{\small II. Institut für Theoretische Physik, Universit{\"a}t Hamburg, Luruper Chaussee 149, D-22761 Hamburg, Germany}}

\date{\today}

\begin{document}

\fontsize{11}{17}\selectfont

\maketitle
\begin{abstract}
The computation of the single-differential top quark-antiquark pair ($\ttbar$)
production cross section at NLO in the fixed-order expansion is examined consistently using
the MSR and $\MSbar$ short-distance top quark mass schemes. A thorough
investigation of the dependence of different regions of the $\ttbar$ invariant mass spectrum on the
renormalization scales $R$ and $\mu_m$ of the MSR mass $\mtMSR(R)$ and $\MSbar$ mass $\mtMSbar(\mu_m)$, respectively, is carried out. We demonstrate that a scale choice of $R\sim 80$~GeV is important for the stability of the cross-section predictions for the low $\ttbar$ invariant mass range, which is important for a reliable extraction of the top quark mass.  
Furthermore, a choice of semi-dynamical renormalization and factorization
scales is preferred. These findings are expected to remain valid once
non-relativistic quasi-bound state effects are included in the low invariant
mass region.
\end{abstract}
\newpage

\section{Introduction}
\label{sec:intro}

The top quark mass $\mt$ is a fundamental parameter of the Standard Model and
has an important role in many predictions, both directly and via higher-order
corrections.  
For instance, together with the values of the strong coupling constant
$\alpS$ and the mass of the Higgs boson, it determines the stability of the
electroweak vacuum~\cite{Bezrukov:2012sa, Degrassi:2012ry,  Alekhin:2012py, Bednyakov:2015sca}.  
Yet, quark masses are formal parameters of the Standard Model QCD Lagrangian and depend on the adopted renormalization scheme. The frequently used top quark pole mass $\mtpole$ is based on the concept of an on-shell observable particle and entails that real and virtual self-energy radiation can be resolved at arbitrarily small energy scales. While the picture of an on-shell top quark allows for consistent cross section field theory computations, it is not physical because the top quark is a colored object and decays. This results in a sizeable linear infrared (IR) sensitivity of $\mtpole$ 
of the order of the QCD scale $\Lambda_{\rm QCD}$~\cite{Bigi:1994em, Beneke:1994sw, Smith:1996xz}\footnote{We note that linear IR sensitivities arise in cross sections whenever cuts on soft radiation are imposed, see e.g.\ Ref.~\cite{FerrarioRavasio:2018ubr}. These are associated to physical nonperturbative corrections in contrast to the pole mass, where the IR sensitivity arises purely from the choice of scheme~\cite{Hoang:2020iah}.} which leads to the pole mass renormalon problem at high orders in perturbation theory.
Quark masses defined in so-called short-distance mass schemes such
as the modified minimal subtraction ($\MSbar$) scheme~\cite{tHooft:1973mfk, PhysRevD.18.3998} 
mass $\mtMSbar(\mu_m)$, or the MSR scheme~\cite{Hoang:2008yj,Hoang:2017suc} mass $\mtMSR(R)$, 
do not have this issue, and their renormalization scales $\mu_m$ and $R$, respectively, act as a finite resolution scale. This means that real and virtual self-energy radiation are treated inclusively for the scales below $\mu_m$ and $R$.
The absence of the ${\cal O}(\Lambda_{\rm QCD})$ renormalon problem, together with the additional freedom to adopt suitable choices for $\mu_m$ and $R$, can be very useful to achieve higher precision concerning the $\mt$ dependence in top mass sensitive observables. In particular, the freedom of scheme and scale choice is important for predictions at lower orders, since it leads to a systematic absorption of sizeable corrections, not originating from the pole mass renormalon, into the quark mass parameter.

A well-known example is the highly \mt-sensitive $\ttbar$ cross section at the production threshold in $e^+e^-$ annihilation, where switching from the pole mass scheme to the threshold mass schemes such as the PS (at the scale of around $25$~GeV) or the 1S mass schemes can minimize the large
$\ttbar$ quasi-bound state corrections at any order, since these mass schemes encode a sizeable fraction of the binding energy corrections $\Delta E\sim \mt\alpS^2$ into the mass value~\cite{Hoang:2000yr}. In the case of the $\ttbar$ quasi-bound state, the scale choice for the PS mass is related to the fact that these $\mt$-sensitive bound state corrections arise from physical dynamical scales of the order of the inverse Bohr radius $\langle 1/r_B\rangle\sim \mt\alpS\sim 25$~GeV, and the MSR mass $\mtMSR(25~\mbox{GeV})$ is a suitable choice as well~\cite{Hoang:2017suc}. Thus, the scale dependence of
$\mtMSbar(\mu_m)$ and $\mtMSR(R)$ allows to properly adapt to these dynamical
scales of the top mass sensitivity for an observable under consideration. The respective renormalization group
equations (RGEs) and matching relations provide the tool to unambiguously relate the \mt{} values extracted at different dynamical scales. This concept is well known for the running strong coupling $\alpS$ and applies to the quark masses as well. 

In this work, the dependence of the invariant mass of the $\ttbar$ pair, $m_\ttbar$, on the MSR mass scale $R$
and the $\MSbar$ mass scale $\mu_m$ is investigated concurrently, for the first time, accounting for QCD corrections. 
Using experimental measurements of $\ttbar$ production at the LHC at $\sqrt{s}=13 \TeV$~\cite{CMS:2019jul}, the next-to-leading order (NLO) prediction of the $m_\ttbar$ differential cross section from Refs.~\cite{Campbell:2010ff, Campbell:2012uf} and the scheme implementation
procedure of Refs.~\cite{Langenfeld:2009wd,Dowling:2013baa}, we demonstrate that the proper scheme choice -- which concerns the proper kind of top mass scheme {\it and} the renormalization scale -- is of key importance and affects the size of higher-order corrections as well as the resulting value of the extracted top quark mass. Our analysis provides in particular a re-investigation of 
the ``indirect'' \mtpole measurement of Ref.~\cite{CMS:2019esx} by the CMS Collaboration which obtained $\mtpole = 170.5 \pm 0.8 \GeV$ based on the same NLO theory prediction. This result is systematically lower than other indirect pole mass measurements~\cite{ATLAS:2015pfy,ATLAS:2019guf,CMS:2022emx} and was reconfirmed by the subsequent analysis in Ref.~\cite{Garzelli:2020fmd}. The latter analysis also extracted the MSR mass at the low scale $R=3$~GeV, $\mtMSR(R=3~\mbox{GeV})$, yielding a low value as well, consistent with Ref.~\cite{CMS:2019esx} after conversion to a common reference scheme.
In the present analysis, we show that using the MSR mass $\mtMSR(R=80~\mbox{GeV})$ would lead to a larger value, compatible with earlier measurements, after conversion to a common reference scheme. We provide general arguments that this particular scale choice leads to a systematic resummation of sizeable higher order QCD corrections that arise in the $\ttbar$ threshold region at the LHC, where the $\ttbar$ pairs are produced in color singlet as well as octet states and sizeable bins of the $m_{\ttbar}$ distribution are used in the analysis. This summation does not take place either in the pole mass scheme or when $\mtMSR(R=3~\mbox{GeV})$ is used, explaining the results obtained in Refs.~\cite{CMS:2019esx,Garzelli:2020fmd}.
The result of our investigation constitutes an important recommendation relevant for upcoming top quark mass measurements from differential $\ttbar$ production cross sections at the LHC. 
We note that the theoretical prediction used in the present analysis does not account for the next-to-next-to-leading order (NNLO) corrections~\cite{Czakon:2015owf,Catani:2019hip} or resummed Coulomb-type quasi-bound state corrections~\cite{Hagiwara:2008df,Kiyo:2008bv,Ju:2020otc}. A coherent theory code that accounts for these results in a way
adequate for the needs of the presented analysis is not yet available in either mass scheme.
We also note that recently the MSR mass has been investigated by the ATLAS collaboration in Ref.~\cite{ATLAS:2021urs} in the context of simulation and top mass calibration studies of the groomed jet mass for boosted top production.

In Sec.~\ref{sec:running_mt_and_xsec}, we review the $\MSbar$ and MSR  top quark mass
schemes and the formulae to implement them, and in
Sec.~\ref{sec:scale_investigations} we carry out a detailed investigation
concerning the best choice of the MSR renormalization scale $R$. In
Sec.~\ref{sec:mtMSR_ext} we quote the results for $\mtMSR(R=1\GeV)$ and higher $R$ values
from the fits to the LHC measurements, demonstrating the impact of the renormalization scale
choice. We close in Sec.~\ref{sec:summary} with a summary and an outlook on future improvements.

\section{Running $\mt$ and the $\ttbar$ pair production cross section at NLO}
\label{sec:running_mt_and_xsec}

In terms of a general mass renormalization scale $\mu_m$, the pole and $\MSbar$ masses are related in perturbative QCD as
\begin{equation}
\label{mtpole-t-mMS}
\mtpole =
\mtMSbar(\mu_m)
\left( 1 + \sum_{n=1} d^{\MSbar}_n(\mu_m) \left(\aspi^{(6)}(\mu_m)\right)^n \right),
\end{equation}
where $\aspi \equiv \alpS/\pi$. Here and everywhere else in this study, we
explicitly indicate by the superscript whether we use the
strong coupling $\alpS^{(5)}$ in the 5-flavor or 
$\alpS^{(6)}$ in the 6-flavor scheme. For the parton distribution functions (PDFs) only
the 5-flavor scheme is employed. All quarks except for the top quark are treated as massless. 
The coefficients $d^{\MSbar}_n(\mu_m)$ in Eq.~\eqref{mtpole-t-mMS} are known up 
to four loops~\cite{Marquard:2015qpa} and the first few orders read~\cite{Gray:1990n, Chetyrkin:1999qi, Melnikov:2000qh}
\begin{align} d^{\MSbar}_1(\mu_m)
&=4/3+L
\, ,
\nonumber\\
d^{\MSbar}_2(\mu_m)
&= 7.1952 + 4.6806 L + 1.4167 L^2
\, ,
 \nonumber \\
d^{\MSbar}_3(\mu_m)
&= 54.161 + 21.776 L + 9.2026 L^2 + 1.7940 L^3
\, ,
\end{align}
where the expansion uses $\alpS^{(6)}$ in the 6-flavor scheme and $L=\log((\mu_m/\overline{m}(\mu_m))^2)$. 
The running of the $\MSbar$ mass is described by the RGE 
\begin{equation}
\mu_m^2\frac{d\mtMSbar(\mu_m)}{d\mu_m^2}
=
- \, \mtMSbar(\mu_m)\sum_{i=0}\gamma_i^m \left(\aspi^{(6)}(\mu)\right)^{i+1}\, ,
\label{MSbar_m_RGE}
\end{equation}
where the anomalous dimensions $\gamma_i^m$ are known to five loops~\cite{Baikov:2014qja,Luthe:2016xec}. 
The first few orders~\cite{Chetyrkin:2000yt, Tarrach:1980up, Tarasov:1982plg, Larin:1993tq, Chetyrkin:1997dh, Vermaseren:1997fq} are given by
\begin{align}
\gamma_0^m &= 1\, ,
\nonumber\\
\gamma_1^m 
&=3.3750\, ,
\nonumber\\
\gamma_2^m
& =4.8387\, ,
\nonumber \\
\gamma_3^m
&= -4.5082\, .
\end{align}
Electroweak corrections (see, e.g.~\cite{Dittmaier:2022maf,Kataev:2022dua}) are not considered.
The RGE in Eq.~\eqref{MSbar_m_RGE} has the solution
\begin{equation}
\mtMSbar(\mu_1)
=
\mtMSbar(\mu_0)
\exp\left\{-2 \sum_{i=0}
         \int_{\mu_0}^{\mu_1} \frac{d\mu}{\mu}
         \gamma_i^m \,\left(\aspi^{(6)}(\mu)\right)^{i+1}
     \right\},
\label{MSbar_m_RGE_sol}
\end{equation}
yielding the $\MSbar$ mass at a scale $\mu_1$ via evolution from the known mass at a reference scale $\mu_0$. 
Here and below we quote relations at ${\cal O}(\alpS^3)$ and
evolution equations at ${\cal O}(\alpS^4)$. 
We have also used these relations in our analysis for determining numerical values for the quark
  masses (and the strong coupling), even though our cross section analysis is
  based on a fixed-order theory description at NLO. Since the mass (and 
  strong coupling) matching relations and RGE equations are well
  convergent series and no subtle cancellations between the different
  ingredients need to be taken care of (which would be the case for the PDFs) this approach is fully consistent and has the
  advantage that the theoretical uncertainties in the numerical values of the 
  masses (and the strong coupling) are eliminated entirely from our analysis. We
  recommend this approach also for future phenomenological analyses. 
For implementing different mass schemes in the analytic expression for the differential $m_\ttbar$ cross sections at NLO, see Eq.~\eqref{diff_NLO_cs} below, 
only the ${\cal O}(\alpS)$ coefficients from Eqs.~\eqref{mtpole-t-mMS} and \eqref{pole2MSR} are used.

The $\MSbar$ mass is by construction a 6-flavor quantity and 
should only be used in observables where the dynamical scale of the top-quark mass sensitivity is of order $\mt$ or larger, 
i.e.\ $\mu_m \gtrsim \mt$. 
The MSR mass is, like the $\MSbar$ mass mass, determined from top-quark self-energy corrections~\cite{Hoang:2017suc,Hoang:2017btd}, but
  designed such that all virtual and off-shell 
  top-quark quantum fluctuations are integrated out in the on-shell limit.\footnote{We are using the natural
    MSR mass definition (MSRn), where virtual top-quark loops are integrated out consistently, see~\cite{Hoang:2017suc}.} 
The MSR mass $\mtMSR(R)$ is therefore a 5-flavor
  quantity and its $R$-dependence properly captures all radiation off the top
  quark that is soft in the top quark rest frame, which is not the case for the
  $\MSbar$ mass. The MSR mass is the proper choice if the dynamical scale of
  the top quark mass sensitivity is below $\mt$, i.e.\ $R\lesssim \mt$.

The pole and MSR masses are related as
\begin{equation}
\mtpole
=
\mtMSR(R)
+ R \sum_{n=1}^{\infty} d^{\textrm{MSR}}_n \left(\aspi^{(5)}(R)\right)^n,
\label{pole2MSR}
\end{equation}
where the coefficients $d^{\textrm{MSR}}_n$ read~\cite{Hoang:2017suc}
\begin{align}
d^{\textrm{MSR}}_1 &= 4/3\, ,
\nonumber\\
d^{\textrm{MSR}}_2 &= 8.1330\, 
\nonumber\\
d^{\textrm{MSR}}_3 &= 71.602\, .
\end{align}
In the limit $R \rightarrow \mtMSbar(\mtMSbar)$, $\mtMSR(R)$ approaches the $\MSbar$ mass 
$\mtMSbar(\mtMSbar)$ and matches on it in analogy to the 5-flavor and 6-flavor strong coupling, see below. 
In contrast to the logarithmic $\mu_m$ evolution of $\mtMSbar(\mu_m)$, the
$R$-evolution of $\mtMSR(R)$ is linear and captures the correct physical
logarithms for observables with $\mt$ dependence, generated at dynamical
scales $R < \mt$, such as resonances, thresholds, and low-energy
endpoints~\cite{Hoang:2021fhn}.  
The mass renormalization constant of the MSR mass only contains the on-shell
self-energy corrections for scales larger than $R$ in contrast to the pole
mass which contains self-energy corrections at all scales. So while the MSR
mass is numerically close to the pole mass for small $R$ at low orders, it is
free of the pole mass renormalon problem. Formally the MSR mass approaches the 
pole mass for $R \rightarrow 0$, but the Landau pole prevents taking this
limit in practice. For small $R$ values in the range of $1$ to $2$~GeV the MSR mass captures the kinematic particle mass interpretation commonly associated of the pole mass. Within perturbative uncertainties at NLO, where we can still
ignore the pole mass renormalon problem, the scheme choice $\mtMSR(R=1\,\mbox{\rm
  GeV})$, or for some other very low value of $R$, is therefore a proxy for the pole mass scheme.  
The matching of the 5-flavor MSR mass to the 6-flavor $\MSbar$ mass at the
scale $R=\mtMSbar(\mtMSbar)$ reads~\cite{Hoang:2017suc} 
\begin{eqnarray}
\mtMSR(\mtMSbar)
&=&
\mtMSbar(\mtMSbar)
\Bigg[1 + 0.10357
           \left( \aspi^{(5)}(\mtMSbar) \right)^2
        + 1.8308
           \left( \aspi^{(5)}(\mtMSbar) \right)^3
\Bigg]\, ,
\label{mMatching}
\end{eqnarray}
and the inverse at the scale $R=\mtMSR(\mtMSR)$ reads~\cite{Hoang:2017suc}
\begin{eqnarray}
\mtMSbar(\mtMSbar)
&=&
\mtMSR \left( \mtMSR \right)
\bigg[ 1 - 0.10357
            \left( \aspi^{(5)}(\mtMSR) \right)^2
 - 1.6927
    \left( \aspi^{(5)}(\mtMSR) \right)^3
\bigg]\, .
\label{mMatchingConverse}
\end{eqnarray}
The matching starts at ${\cal O}(\alpS^2)$, where virtual top quark loops
first appear. In the matching relations in Eqs.~\eqref{mMatching} and \eqref{mMatchingConverse} we have indicated the 5-flavor scheme for the strong coupling. At the order shown, the coefficients are identical to the ones in the 6-flavor scheme. 
These relations are in close analogy to the corresponding strong coupling matching relation which reads
\begin{equation}
\aspi^{(6)}(\mtMSbar) 
= \aspi^{(5)}(\mtMSbar)\bigg[1 - 0.15278
\,\left(\aspi^{(5)}(\mtMSbar)\right)^{2}
-\,0.54881 
\,\left(\aspi^{(5)}(\mtMSbar)\right)^{3} \bigg].
\label{mMatchingalphasnew}
\end{equation}
Note, the corrections to the matching relation shown in the brackets of Eq.~(\ref{mMatchingalphasnew}) are known to ${\cal O}(\alpha_s^4)$~\cite{Chetyrkin:2005ia,Schroder:2005hy}.
The MSR mass at an arbitrary scale $R$ is then obtained from a given $\MSbar$
mass, applying Eq.~\eqref{mMatching}, and evolving the scale $R$ from
$\mtMSbar(\mtMSbar)$ to the desired value by solving the RGE 
\begin{equation}
R\frac{d}{dR}\mtMSR(R)
=
- R \sum_n
  \gamma_n^R \left(\aspi^{(5)}(R)\right)^{n+1},
\label{m_R_RGE}
\end{equation}
where the anomalous dimensions $\gamma_n^R$ are given by~\cite{Hoang:2017btd}
\begin{align}
\gamma^R_0 &=  
4/3\, 
\nonumber \\ 
\gamma^R_1 
&= 3.0219\, ,
\nonumber \\
\gamma^R_2 
&= 2.8047\, ,
\nonumber \\
\gamma^R_3 
&= -73.257 \, .
\end{align}
The solution of Eq.~\eqref{m_R_RGE} yields
\begin{equation}
\mtMSR(\mtMSbar) - \mtMSR(R)
=
- \sum_{n=0}
\gamma_n^R
\int_{R}^{\mtMSbar} dR' \left(\aspi^{(5)}(R')\right)^{n+1}
+\mathcal{O}\left(\aspi^4\right)
\equiv
\Delta m\, ,
\label{mRGEsol}
\end{equation}
so that the MSR mass at $R$ is obtained as $\mtMSR(R) = \mtMSR(\mtMSbar) - \Delta m$. 
  As far as QCD corrections are concerned, the
formulae above allow to relate MSR and $\MSbar$ top quark mass values at any
(perturbative) scale with a precision of better than $20$~MeV. The REvolver
library~\cite{Hoang:2021fhn} provides this functionality in a user-friendly software package.

In the present work, the \MCFM program (version 6.8)~\cite{Campbell:2010ff,
  Campbell:2012uf} is extended to include the implementation of the MSR scheme
in the computation of the hadronic $\ttbar$ production cross section for single-differential kinematics. 
Based on the procedure presented in Refs.~\cite{Langenfeld:2009wd,Dowling:2013baa}, 
the
$\ttbar$ production cross section differential with respect to an observable $X$ at NLO reads 
\begin{align}
\frac{d\sigma}{dX}
&=  \left( \aspi(\mur) \right)^2
   \frac{d\sigma^{(0)}}{dX}\big(m,\mur,\muf\big)
 + \left( \aspi(\mur) \right)^3
   \frac{d\sigma^{(1)}}{dX}\big(m,\mur,\muf\big)
\nonumber\\
&+ \left( \aspi(\mur) \right)^3
   \,\tilde{R}\, d_1\,
   \frac{d}{d \mt}\left(\frac{d\sigma^{(0)}(\mt,\mur,\muf)}{dX}\right)
                 \bigg|_{\mt = m}\, ,
\label{diff_NLO_cs}
\end{align}
where $\sigma^{(0)}$ is the leading order (LO) and $\sigma^{(1)}$ the NLO
cross section in the pole mass scheme. At NLO, the derivative term (the third
summand in Eq.~\eqref{diff_NLO_cs}) implements the $\MSbar$ or MSR top quark
mass schemes. 
In the present work, the observable of interest is the invariant mass of the
$\ttbar$ system, and $X = m_\ttbar$. 
In particular, we have the following set of parameters in Eq.~\eqref{diff_NLO_cs})
\begin{equation}
\label{eq:parameters}
\left(\aspi(\mur),m, d_1, \tilde{R}\right)
= 
\begin{cases}
\left(\aspi^{(5)}(\mur),\mtMSR(R),~d^{\textrm{MSR}}_1,~R\right)\, ,
&\textrm{$R<\mtMSbar(\mtMSbar)$ (MSR regime)}\, ,\\
\left(\aspi^{(5)}(\mur),\mtMSbar(\mu_m),~d^{\MSbar}_1(\mu_m),~\mtMSbar(\mu_m)\right)\, ,
&\textrm{$\mu_m>\mtMSbar(\mtMSbar)$ ($\MSbar$ regime)}\, .
\end{cases}
\end{equation}
It is important to note that the choice of the renormalization and factorization scales $\mur$ and $\muf$ is independent of the mass renormalization scales $R$ or $\mu_m$ in this implementation.
We emphasize that it is essential that the mass scheme correction proportional to $d_1$ is consistently used at the renormalization scale $\mur$, which yields logarithms $\ln(R/\mur)$ or $\ln(\mu_m/\mur)$ beyond NLO to consistently cancel the pole mass renormalon. 
Since MCFM is based on renormalization with 5 dynamical flavors, one has to consistently expand $\aspi^{(6)}(\mu_r)$ for the $\MSbar$ top mass scheme corrections of Eq.~\eqref{mtpole-t-mMS} in powers of $\aspi^{(5)}(\mu_r)$ in the cross section formula of Eq.~\eqref{diff_NLO_cs}. At NLO this leads to 
Eq.~\eqref{eq:parameters}.

We note that the fixed-order perturbative corrections for the differential cross section in the pole mass scheme are known at NNLO  accuracy in QCD~\cite{Czakon:2015owf,Catani:2019hip} 
and at NLO in the electroweak theory~\cite{Kuhn:2006vh,Czakon:2017wor}.
In addition, the $\overline{\rm MS}$ mass scheme at NNLO has been studied in Ref.~\cite{Catani:2020tko}. 
The conversion of the mass renormalization scheme from the pole mass to the $\overline{\rm MS}$ or the MSR mass beyond NLO accuracy in QCD (and LO for electroweak effects as presented here) needs to be performed numerically 
and requires theory predictions for differential cross sections
with the pole mass at NNLO accuracy for a large array of pole mass
values (typically in a range $150 \GeV < m < 180 \GeV$)). 
The required NNLO computations (including scale variations and PDF uncertainties) are numerically demanding and such results are currently not readily available in the literature.

Non-relativistic quasi-bound state QCD corrections are important for the region $m_\ttbar\sim 340$-$360$~GeV, 
where the strongest top quark mass sensitivity arises in the $m_\ttbar$ distribution.
In this threshold region the produced top
quarks attain small non-relativistic velocities $v \ll 1$ in the $\ttbar$ 
center-of-mass frame, and the dynamics of
the $\ttbar$ system are hence governed by the mass $\mt$, the relative
momentum $\mt v$, and the kinetic energy $\mt v^2$ of the top quark. Since $\mt
\gg \mt v \gg \mt v^2$, the appearance of ratios involving the masses, momenta
and kinetic energy of the top quark renders the standard fixed-order expansion
in powers of $\alpS$ unreliable in this $m_\ttbar$ range. The most pronounced quasi-bound state effects arise from the Coulomb corrections due to the exchange of gluons between the produced $\PQt$ and $\PAQt$ yielding a dependence of the
prediction on the ratio $\mt/(\mt v)$. This leads to a singular
$(\alpS/v)^n$ behavior in the fixed-order perturbative QCD correction at
$n$-loops~\cite{Hoang:1999zc}.
The quasi-bound state effects have been considered in
Refs.~\cite{Hagiwara:2008df,Kiyo:2008bv}, and more recently again
in~\cite{Ju:2020otc}. For $\ttbar$ pairs in a color octet state, the effects of soft gluon exchanges with other parts of the hard production process can yield further significant corrections, which are currently unknown. The available predictions do not provide an adequate
description of the lowest $m_\ttbar$ bin in the region between $300$~GeV and
the quasi-bound state region around $350$~GeV, 
where the imaginary energy approach and the use of the optical theorem~\cite{Hoang:2000yr} predict a sizeable and unphysical finite $\ttbar$
production rate, see the results shown in Ref.~\cite{Ju:2020otc}. 
In this region the differential cross section depends on the experimental cuts on the
top and antitop quark decay products~\cite{Hoang:2010gu,Beneke:2017rdn}, which
complicates the theoretical prediction as well as the experimental analysis,
but any sensible choice of cuts leads to a strongly suppressed rate for
$m_\ttbar$ close to $300$~GeV. 
This latter aspect is actually better described
by the fixed-order predictions for stable top quarks where the rate vanishes
identically for $m_\ttbar<2\mt$ (for a correct top mass scheme choice as discussed below). 
Furthermore, a systematic treatment of the intermediate region, where the
non-relativistic and relativistic calculations need to be matched, 
is currently not available with a reliable matching error estimate.\footnote{Such a treatment is available only for top quark production in $e^+e^-$ annihilation, see Ref.~\cite{Bach:2017ggt}.} 
We also mention that for the electroweak corrections different scheme choices
for the $\MSbar$ mass are available related to the definition of the vacuum
expectation value~\cite{Dittmaier:2022maf,Kataev:2022dua}. Their effects concerning the
MSR mass and their impact on the use of different mass schemes in experimental
observables are unknown. 
Overall, there is currently no complete and reliable theory prediction 
for the low $m_\ttbar$ distribution available for experimental analysis.  
For the study of the $\ttbar$  differential cross section as a function of $m_\ttbar$ and its dependence on the MSR mass scale $R$, 
the NLO fixed order prediction for stable top quarks based on the \MCFM
program is appropriate, since it properly describes the generic size of
subleading QCD corrections and vanishes for $m_\ttbar<2\mt$. 
For a reliable measurement of the MSR top quark mass, however, a more complete
code including the features mentioned above has to be made available.

\section{First investigation of the $R$ scale dependence}
\label{sec:scale_investigations}

In this section we examine the dependence of the $m_\ttbar$ distribution in
different representative bins in the range between $300$ and $700\GeV$ on the scales $\mur$, $\muf$, and $R$ in the MSR
mass scheme as well as $\mu_m$ in the \MSbar{} scheme using as input the results of the ABMP16 PDF fit at NLO~\cite{Alekhin:2018pai}
with
$\alpS^{(5)}(m_\mathrm{Z})=0.11905$ at $m_\mathrm{Z}=91.19\GeV$. 
For the $\MSbar$ mass the value $\mtMSbar(\mtMSbar)=160.68$~GeV has been chosen close to the fit of Ref.~\cite{Alekhin:2017kpj}.
The latter value corresponds to MSR masses at $R=1\GeV$ and $R=80\GeV$ of $\mtMSR(1\GeV)=170.48$~GeV and $\mtMSR(80\GeV)=164.98$~GeV, respectively. 

\begin{figure}[b!]
\centering
\includegraphics[width=0.495\textwidth,trim={0mm 0mm 0mm 11mm},clip]{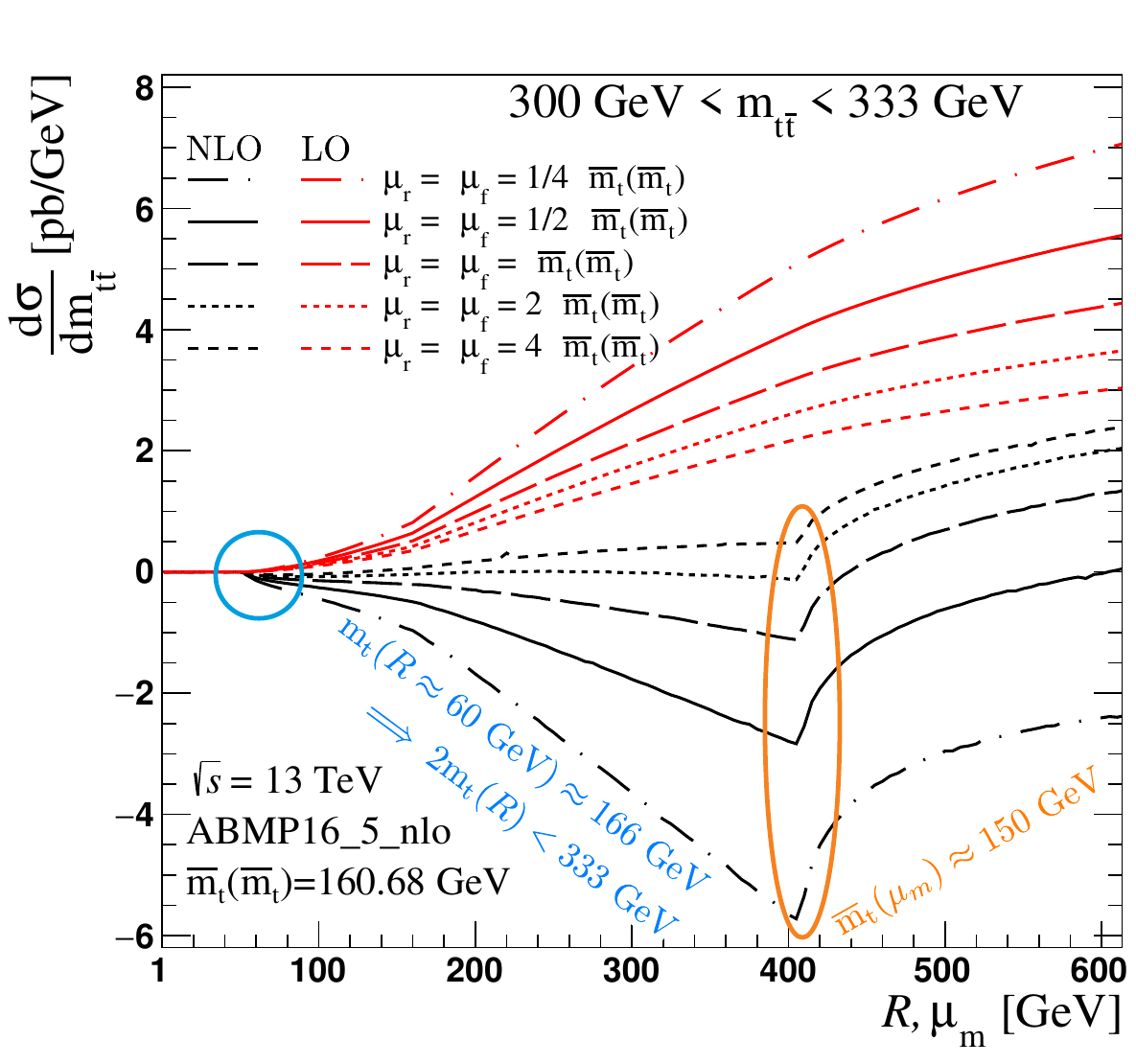}
\caption{The $m_\ttbar\in [300, 333]\GeV$ range of the $m_\ttbar$
  distribution. There is no $\ttbar$ production at $R\lesssim 60\GeV$, but the
  region above it suffers from the lack of Coulomb corrections. The
  discontinuity at $\mu_m \gtrsim 410\GeV$ is due to the $\ttbar$ production
  threshold becoming artificially low, and such high values of the scale
  $\mu_m$ should be avoided.} 
\label{xsec_300-333_GeV}
\end{figure}
In Fig.~\ref{xsec_300-333_GeV}, the cross section for the bin $m_\ttbar \in
[300,333] \GeV$, i.e.\ the region below the $\ttbar$ production threshold, is
shown for different scale choices at LO and NLO. The cross section is zero for
$R< 60 \GeV$, which corresponds to $2\mtMSR(R)>333 \GeV$.
Non-zero contributions to the cross section in the $m_\ttbar\in[300, 333]\GeV$
range appear only at large values of $R$ or when using the \MSbar{} mass,
which correspond to smaller values of $\mtMSR(R)$ or $\mtMSbar(\mu_m)$.  
The LO contribution to the cross section is zero or positive throughout the probed range of $R$ and $\mu_m$. 
At NLO, however, the quick decrease of the derivative terms in Eq.~\eqref{diff_NLO_cs}
in comparison to the increase of the positive contributions would lead to
unphysical negative values of the NLO cross section in this kinematic range, as was also pointed out in Ref.~\cite{Catani:2020tko}, where the \MSbar{} mass scheme was examined.  

Since $\ttbar$ production in the range $m_\ttbar \in [300,333] \GeV$ 
is impossible, the results in Fig.~\ref{xsec_300-333_GeV} 
also show that $R$ values above $80$~GeV must be avoided. 
This also implies that the $\MSbar$ mass cannot be used if the $\ttbar$ cross section in this $m_\ttbar$ range is included in the experimental analysis. 
This conclusion holds even in the presence of quasi-bound state effects, 
since these provide a more precise prediction of the $\ttbar$ production threshold, which
is, however, located at $m_\ttbar$ values above $333$~GeV. 
A further feature of the $m_\ttbar\in[300, 333]\GeV$ range, shown in
Fig.~\ref{xsec_300-333_GeV}, is the rapid increase of the cross section at
$\mu_m \gtrsim 410\GeV$. This occurs when $\mtMSbar(\mu_m)$ is so small, such that LO
$\ttbar$ production is even possible below $300\GeV$.

\begin{figure}[t!]
\centering
\includegraphics[width=0.495\textwidth,trim={0mm 0mm 0mm 0mm},clip]{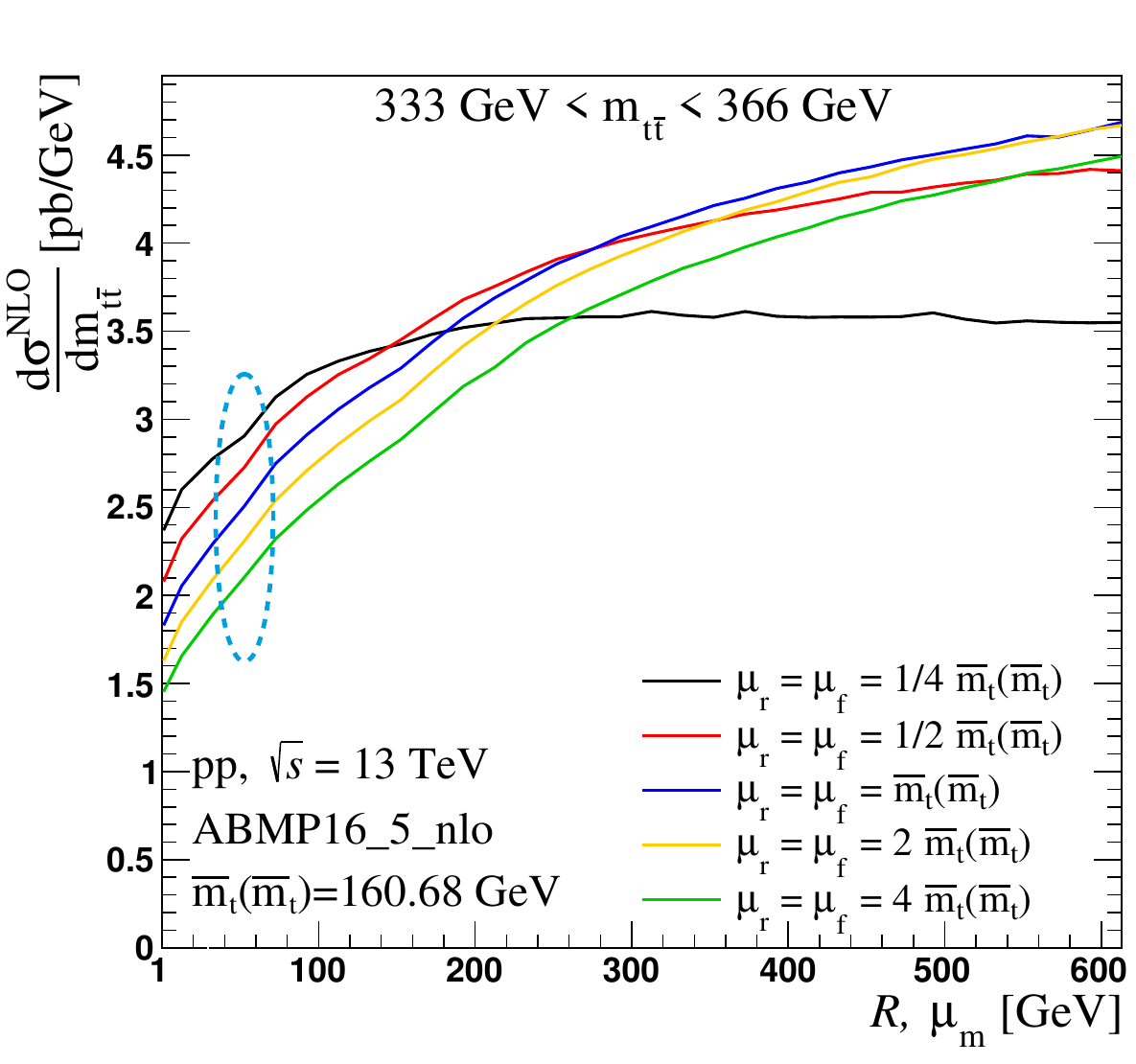}
\includegraphics[width=0.495\textwidth,trim={0mm 0mm 0mm 11mm},clip]{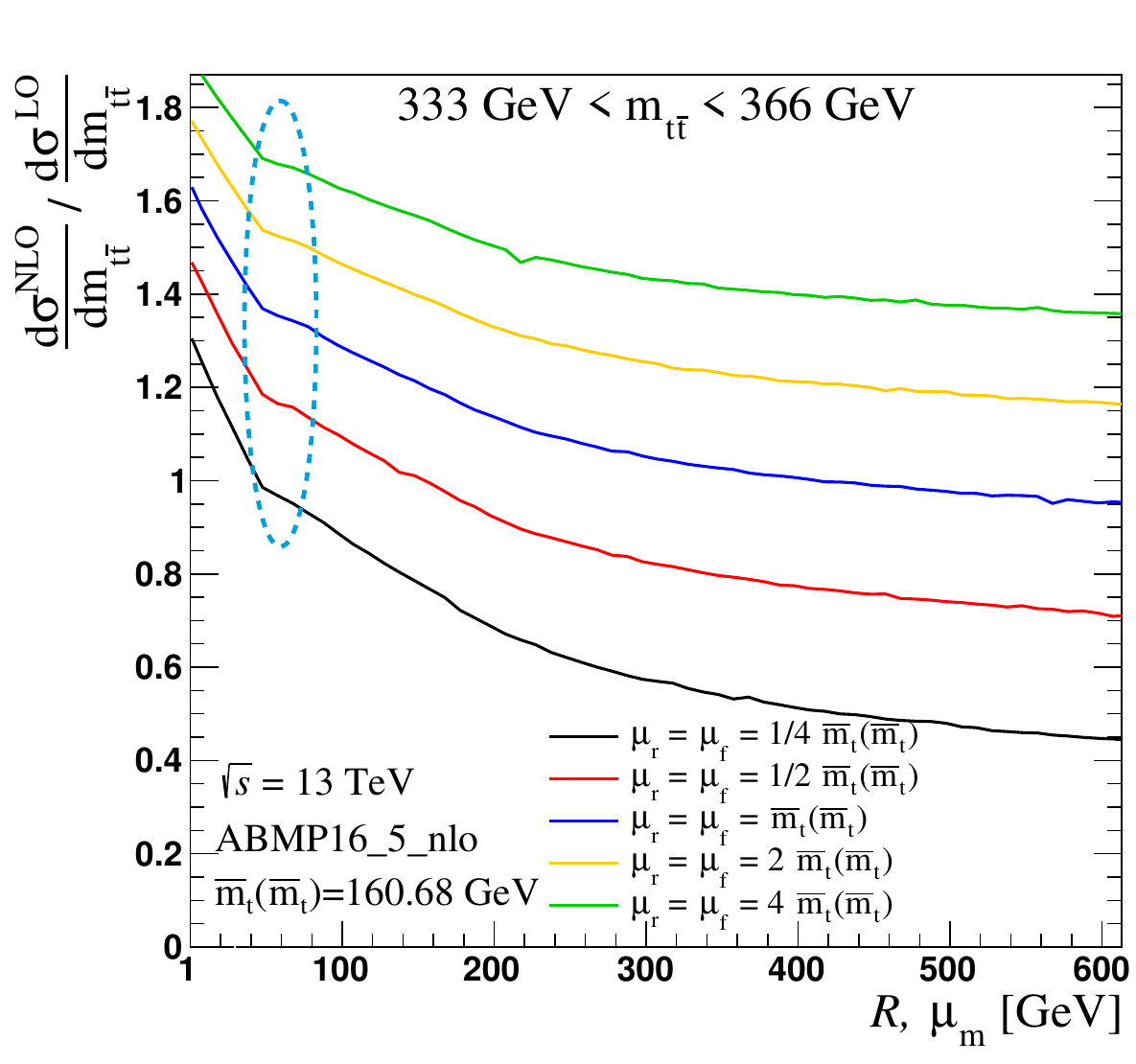}
\caption{The NLO cross section (left) and the ratio of the LO and NLO cross
  sections (right) for $m_\ttbar \in [333, 366]\GeV$. The transition from a
  region suffering from the missing Coulomb corrections to a more stable
  region where the threshold effects become less important is seen at $R
  \gtrsim 60\GeV$ (dashed blue). Further, predictions obtained using small
  values of $\mur,~\muf$ are observed to stabilize the prediction quickly as a
  function of $R$ or $\mu_m$.} 
\label{ratio_333-366_GeV}
\end{figure}

In Fig.~\ref{ratio_333-366_GeV}, 
the cross section for the bin $m_\ttbar \in [333,366] \GeV$, 
i.e. the region where the $\ttbar$ production threshold is located, 
is shown as a function of $R$ and $\mu_m$ at NLO  in the left panel. 
The right panel displays the relative size of the NLO corrections with respect to the LO description. 
Here, the quasi-bound state effects already contained in the NLO prediction are sizeable and our NLO result only provides a qualitative description. 
Similar as in the lowest bin, we observe a quite strong dependence on the mass renormalization scale. 
We see that for very small values of $R$ the size of the NLO correction increases significantly, particularly for large $\mur$ and $\muf$ values, making the use of fixed-order perturbation theory unreliable for these choices. 
This shows that the impact of the higher-order QCD corrections, including the quasi-bound state corrections, is particularly sizeable and essentially maximized in the pole mass scheme. This is closely mimicked by the result for $R=1\GeV$. 

On the other hand, with increasing $R$, the impact of the NLO corrections decreases substantially. 
This is illustrated in the right panel of Fig.~\ref{ratio_333-366_GeV}, where the ratio of cross sections at NLO to that at LO is shown to be closer to unity. This should be interpreted as the NLO corrections being small at these $R$. 
Particularly, with $R$ in the range $60\GeV$ to $80\GeV$, the cross section remains robust under variations of all the scales: it changes only little as a function of $R$, while the differences between the curves corresponding to smaller or larger central values for $\mur$ and $\muf$ remain small. 
This is not accidental, but expected from the fact that the smaller value of the MSR mass at larger $R$ values accounts for the fact that the mass of the $\ttbar$ system is reduced on average by the Coulomb-binding effects. Therefore, with $R$ in the range $60\GeV$ to $80\GeV$, also the impact of the (missing) Coulomb corrections can be expected to be moderate and in particular much smaller than they would be for very small values of $R$, which serve as a proxy for the pole mass scheme. 
It is also observed that setting $\mur$ and $\muf$ to values below the top quark mass further diminishes the size of the NLO corrections. 
The reason is that for this particular $R$-range and chosen $\mur$ and $\muf$, 
$\mtMSR (R)$ captures a sizable part of the non-relativistic bound-state dynamics relevant in the region $m_\ttbar \in [333, 366] \GeV$. 
In other words, a significant part of the binding effects is already absorbed into the mass value, so that the fixed-order prediction at NLO in the MSR scheme provides a more precise description of the process as compared to the pole mass scheme.
Considering the observations in the $m_\ttbar$ range of $[300,366]$~GeV (Fig.~\ref{xsec_300-333_GeV} and Fig.~\ref{ratio_333-366_GeV}), it can be concluded that the most stable predictions are obtained for $R$ in the range of $60$ to $80 \GeV$. 
Out of this range, the highest choice for the scale, $R=80\GeV$ is preferred as the default choice, to avoid entering the regime of $R<60 \GeV$ while allowing the examination of the $R$ scale variation uncertainties, see Sec.~\ref{sec:mtMSR_ext}.
From the experience gained in studies of the $\ttbar$ threshold scan at $e^+e^-$-colliders~\cite{Hoang:2000yr} one may expect that a MSR mass scale $R\sim 25\GeV$ should be more appropriate. However, due to the integration over the $m_\ttbar$ range, the values of $R$ and $\mur$ need to be larger, than those used to describe the peak of the bound-state resonance, corresponding to a much narrower $m_\ttbar$ distribution.

\begin{figure}[t!]
\centering
\includegraphics[width=0.495\textwidth,trim={0mm 0mm 0mm 0mm},clip]{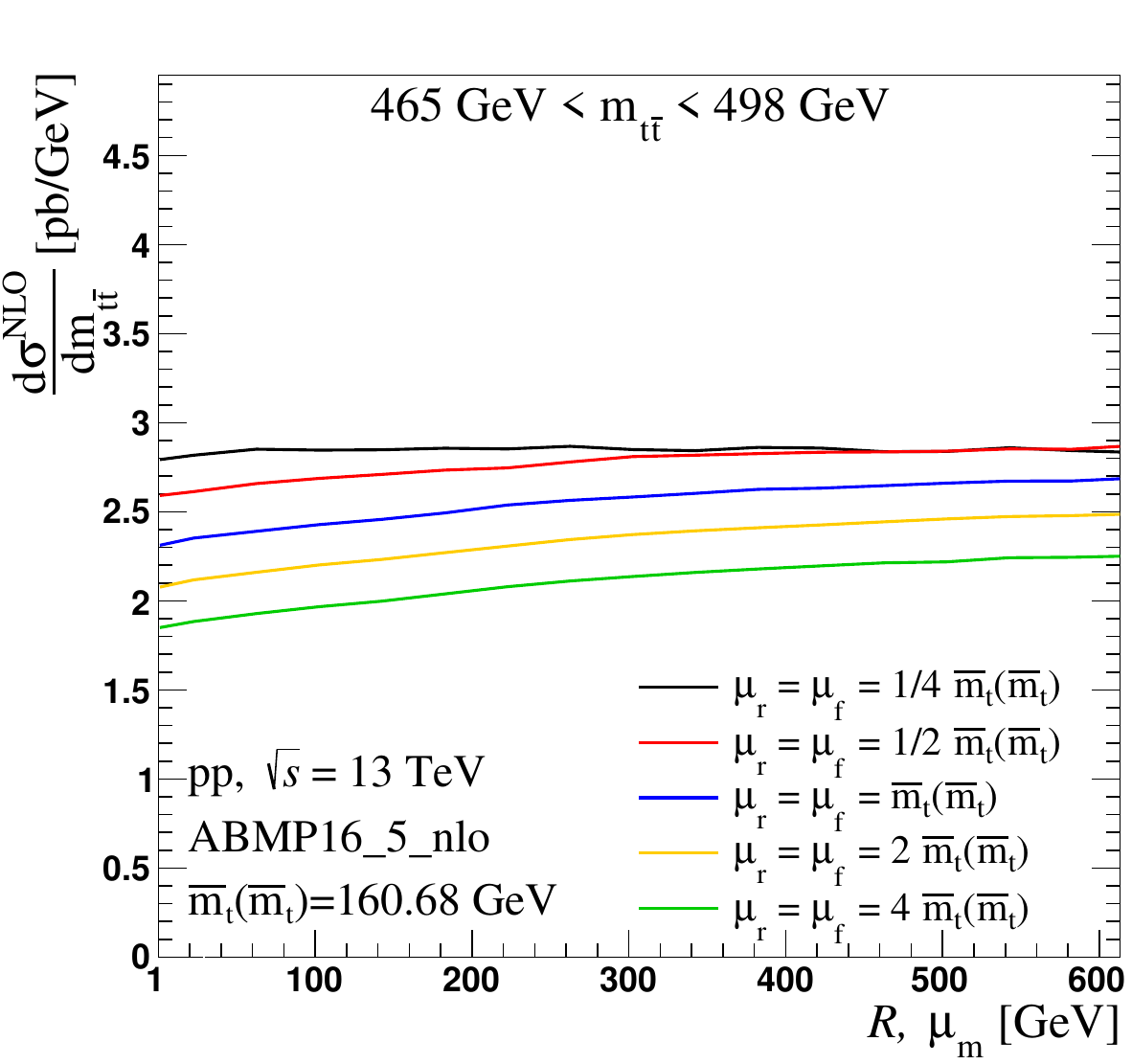}
\includegraphics[width=0.495\textwidth,trim={0mm 0mm 0mm 11mm},clip]{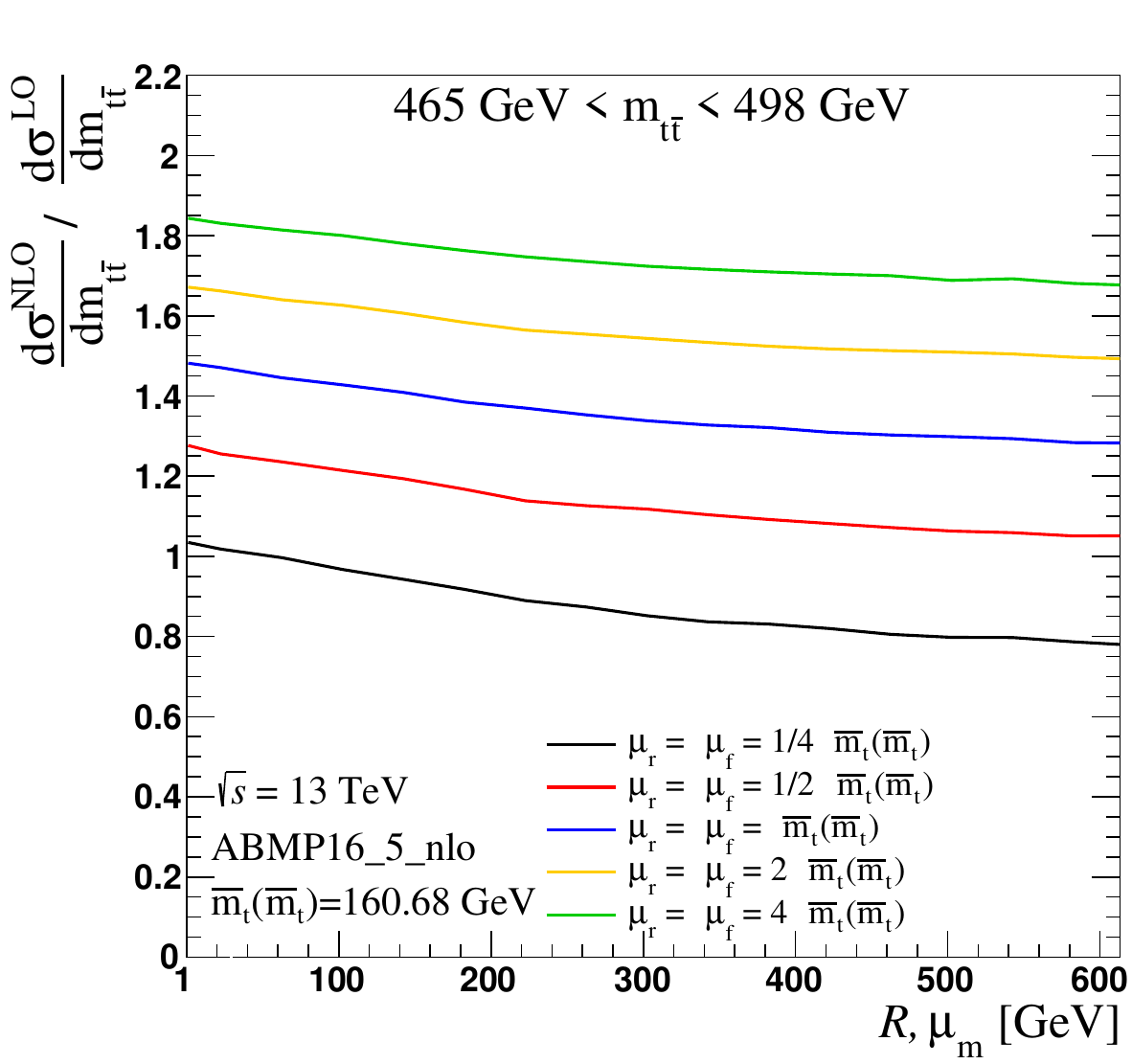}
\caption{The NLO cross section (left) and the ratio of the LO and NLO cross
  sections (right) for $m_\ttbar \in [465, 498]\GeV$.} 
\label{ratio_465-498_GeV}
\includegraphics[width=0.495\textwidth,trim={0mm 0mm 0mm 0mm},clip]{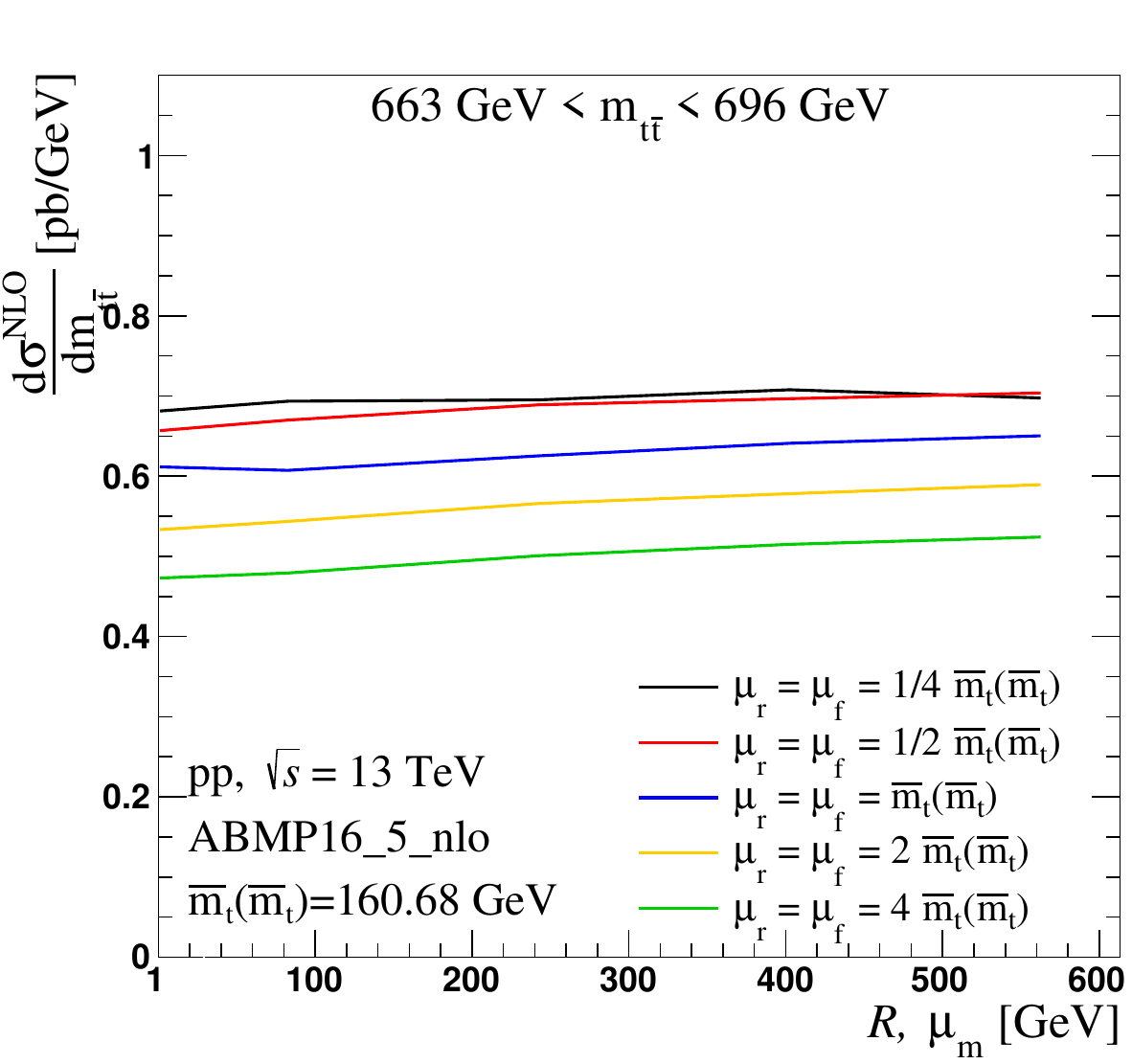}
\includegraphics[width=0.495\textwidth,trim={0mm 0mm 0mm 11mm},clip]{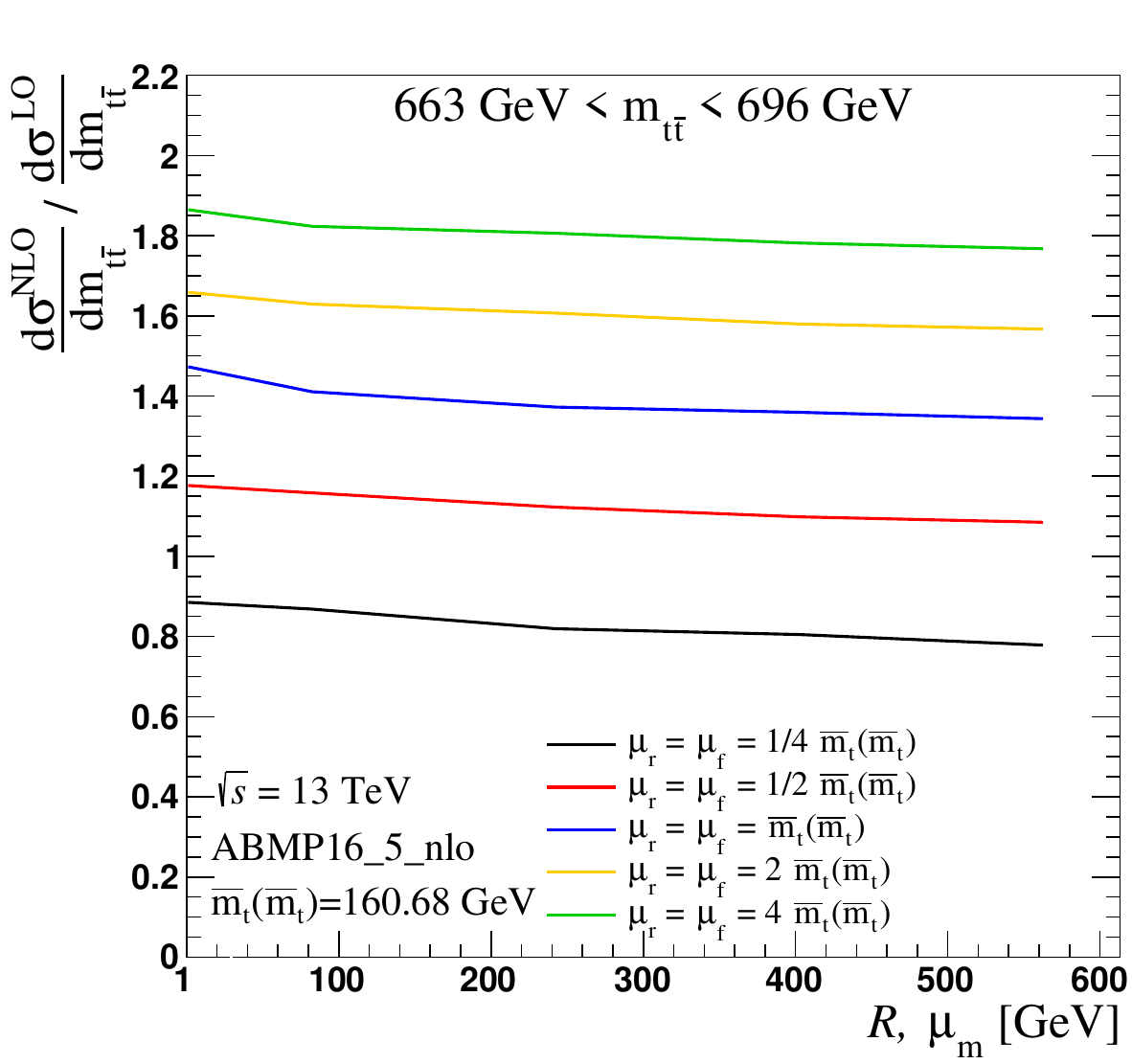}
\caption{
Same as Fig.~\ref{ratio_465-498_GeV} for the bin $m_\ttbar \in [663, 696]\GeV$.}
\label{ratio_663-696_GeV}
\end{figure}
At this point it is also instructive to examine $m_\ttbar$ far above
threshold. In Figs.~\ref{ratio_465-498_GeV} and \ref{ratio_663-696_GeV}, the results for $m_\ttbar \in [465,498] \GeV$ and $m_\ttbar \in
[663,696] \GeV$, respectively, are shown. Here the NLO predictions provide an appropriate 
theoretical description. In contrast to the low $m_\ttbar$ bins discussed above, the mass
renormalization scale behavior is very smooth. This is
partly related to the much smaller top quark mass sensitivity, but also means that
none of the top quark mass schemes (and values for $R$ or $\mu_m$) provide any advantage
concerning capturing essential QCD corrections. 
Here, only the choices of the scales $\mur$ and $\mu_f$ are essential for the prediction showing a preference for values of around $\mt$. 
This observation applies also to other invariant mass bins covering large $m_\ttbar$ values, see Ref.~\cite{Maekelae:482499}.

Overall, our examination suggests that the MSR top quark mass 
$\mtMSR(R)$ and the choice for the
central value of $R=80$~GeV provide the most reliable theoretical predictions for
all $m_\ttbar$ bins. For the scales $\mur$
and $\muf$ the central values $\mtMSbar(\mtMSbar)$ and, in particular $\mtMSbar(\mtMSbar)/2$ for the $m_{\ttbar}$ range containing the $\ttbar$ threshold, are adequate choices. 
We note that these findings are also in line with the optimal scale choices for the total cross section for $\ttbar$ hadro-production, 
when using the top quark mass in the $\MSbar$ scheme. In this case, central values for $\mur$ and $\muf$ 
of the order $\mtMSbar(\mtMSbar)/2 \approx 80\GeV$ are in the region of fastest apparent convergence 
considering perturbative QCD corrections through NNLO and also minimize the scale sensitivity of the total cross section~\cite{Dowling:2013baa}.  
Settings for PDF factorization scale $\mu_f$ different from $\mu_r$ have been explored in Refs~\cite{Catani:2020tko,Garzelli:2020fmd}, corroborating these findings. 
On the other hand, for the total cross section with the top quarks in the pole mass 
scheme, which is well modeled by the MSR scheme mass $\mtMSR(1\GeV)$, the preferred central values for $\mur$ and $\muf$, which minimize scale sensitivity and optimize perturbative convergence through NNLO, are of the order $\mtpole/4 \approx 45\GeV$, see e.g.\ Ref.~\cite{Dowling:2013baa}.
This is also visible in the ratio plots on the right in Figs.~\ref{ratio_333-366_GeV}--\ref{ratio_663-696_GeV}. 
In the following, we demonstrate the impact of the mass scheme and the scale setting on the value of the top quark mass obtained in fits to the experimental data of Ref.~\cite{CMS:2019jul}.

\section{Extraction of the top quark MSR mass}
\label{sec:mtMSR_ext}

The MSR mass $\mtMSR(R)$ is extracted from the differential $\ttbar$
production cross section measured by the CMS Collaboration in $\pp$ collisions
at the LHC at $\sqrt{s}=13\TeV$, corresponding to an integrated luminosity of
$35.9 \fbinv$~\cite{CMS:2019jul}. The $\ttbar$ cross section is measured as a function of $m_\ttbar$ in the ranges: 
$m_\ttbar < 420 \GeV$,
$m_\ttbar \in [420,550] \GeV$,
$m_\ttbar \in [550,810] \GeV$
and $m_\ttbar > 810 \GeV$.

The theoretical predictions are obtained using the ABMP16 5-flavor PDF set~\cite{Alekhin:2017kpj} at NLO. 
According to the preferred MSR mass scale settings described in the previous section, the
initial value of the scale $R$ is set to $80\GeV$ in Eq.~\eqref{diff_NLO_cs}, and the cross section is
calculated for a range of assumed values of $\mtMSR(80\GeV)$. The function 
\begin{equation}
\chi^2 = \sum_{i,j} (\sigma^{\textrm{exp}}_i - \sigma^{\textrm{th}}_i)
                    C^{-1}_{ij}
                    (\sigma^{\textrm{exp}}_j - \sigma^{\textrm{th}}_j),
\label{chi2}
\end{equation}
is computed for each $\mtMSR(80\GeV)$. The indices $i,j$ in Eq.~\eqref{chi2}
run over the bins of the $m_\ttbar$ distribution, while
$\sigma^{\textrm{exp}}_i$ are the experimental data and
$\sigma^{\textrm{th}}_i$ the theoretical predictions. The inverse covariance
matrix $C^{-1}_{ij}$ provided in Ref.~\cite{CMS:2019jul} is used. 

\begin{figure}[ht]
\centering
\includegraphics[width=0.495\textwidth]{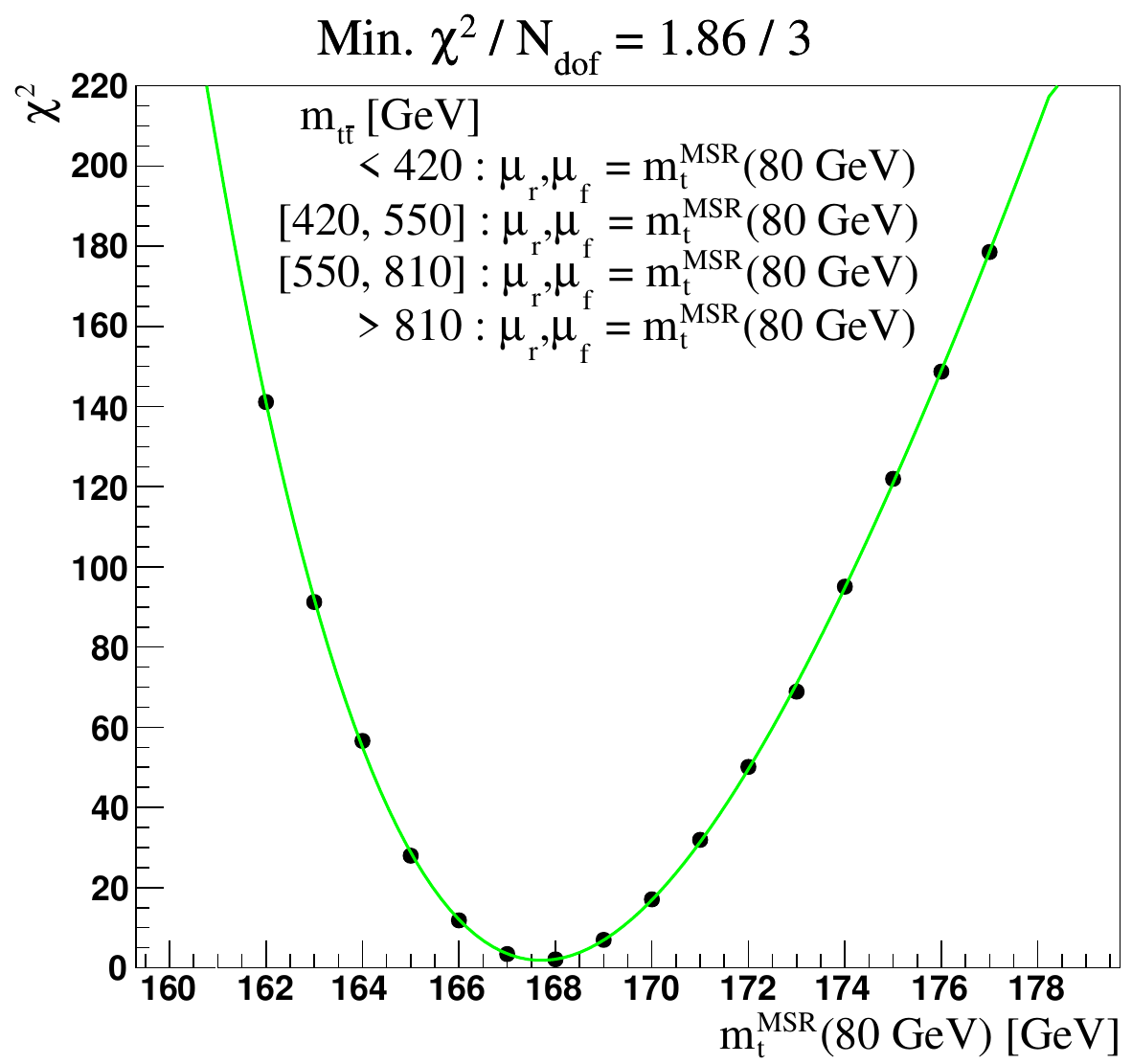}
\includegraphics[width=0.495\textwidth]{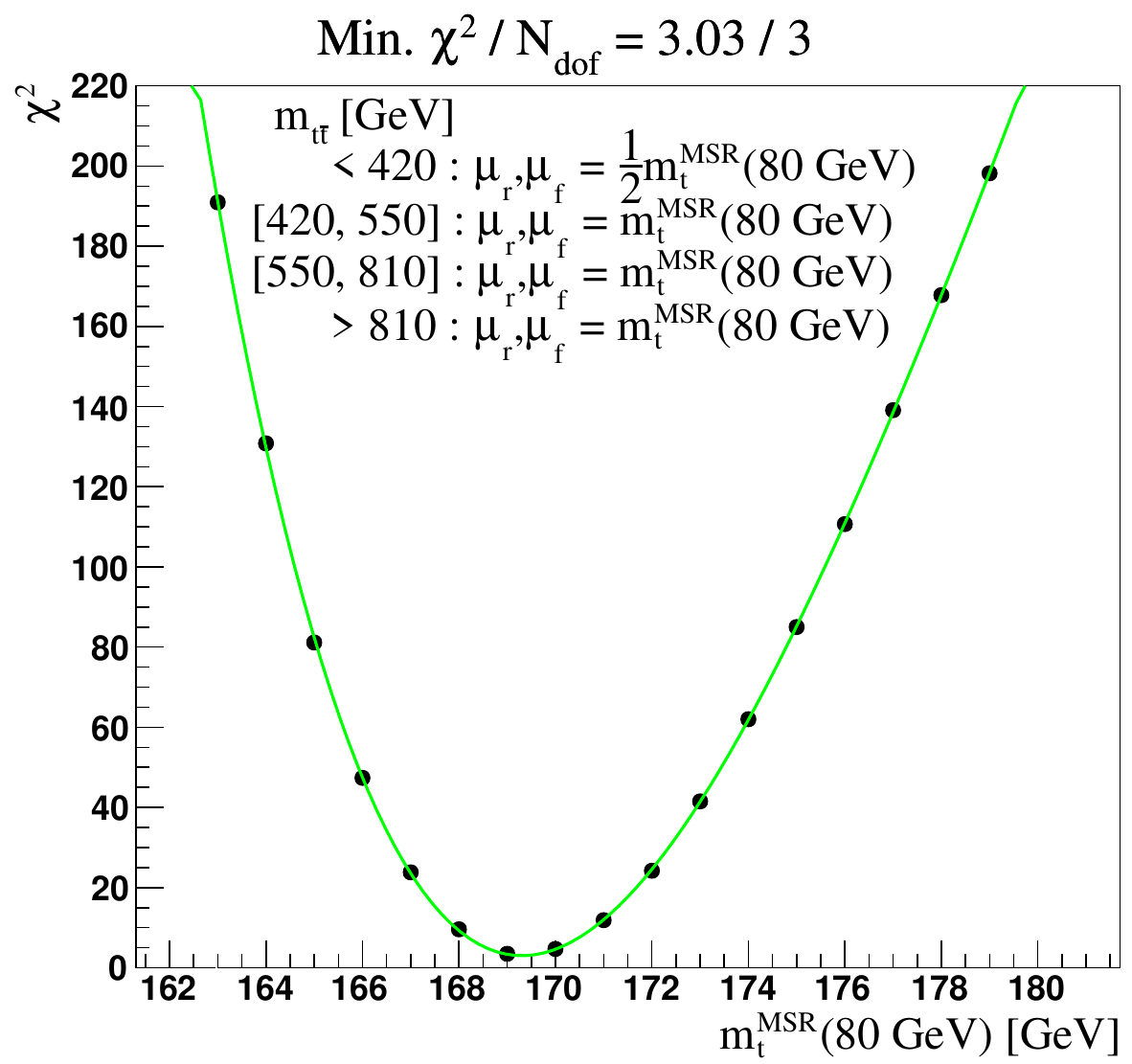}
\caption{A 4th order polynomial fitted to the $\chi^2$ resulting from
  comparing the experimental data to theory predictions assuming different
  values of $\mtMSR(80\GeV)$. The scales $\mur$ and $\muf$ are set to
  $\mtMSR(80\GeV)$ considering the whole $m_\ttbar$ distribution (left), or to
  $\mtMSR(80\GeV)/2$ for $m_\ttbar < 420 \GeV$ and to $\mtMSR(80\GeV)$ for the
  remainder (right). The number of degrees of freedom in the fits is denoted
  by $\mathrm{N}_\mathrm{dof}$.}
\label{mtFit_chi2}
\end{figure}

The scales $\mur$ and $\muf$ are set to $\mtMSR(80\GeV)$ for all 4 bins of the
$m_\ttbar$ distribution or, alternatively, to $\mtMSR(80\GeV)/2$ for 
$m_\ttbar < 420 \GeV$, to stabilize the prediction against the missing
quasi-bound state corrections, and to $\mtMSR(80\GeV)$ for the remainder.  
Fig.~\ref{mtFit_chi2} shows a 4th order polynomial fit to the $\chi^2$ values
resulting from each configuration. 

The fit uncertainties are obtained via the $\Delta\chi^2=1$ tolerance
criterion, while the $\mur$ and $\muf$ scale uncertainties are evaluated by
varying their central values in each bin up and down by a factor of 2,
avoiding the cases where one scale is multiplied by 1/2 and the other by 2,
and constructing an envelope. 
For comparison with previous analyses, the extracted values of
$\mtMSR(80\GeV)$ are evolved to the reference scales $R$ of 1 and 3\GeV. Note
that determining $\mtMSR(1 \GeV)$ requires evaluating
$\alpS(1\GeV)$ rather close to the Landau pole, which is expected to lead to an 
increased perturbative uncertainty in the MSR mass at $R=1\GeV$ due to missing higher order corrections. 
Reporting the mass value also at $R=3\GeV$ thus ensures the stability of the result, 
and the use of reference scales $R>1\GeV$ will become increasingly important in future
extractions of $\mtMSR(R)$. Furthermore, the results are translated into the
standard $\MSbar$ mass $\mtMSbar(\mtMSbar)$ by iteratively 
finding $\mtMSR(\mtMSR)$ via the condition $R=\mtMSR(R)$, and applying the
matching formula in Eq.~\eqref{mMatchingConverse} up to $\mathcal{O}(\aspi^3)$. 
The uncertainty related to the initial choice of $R$ is assessed by repeating the fits
at $R=60\GeV$ and $R=100\GeV$, and the difference in the resulting masses at
the reference scales to the respective values obtained in the $R=80\GeV$ fit
is taken as the $R$ scale uncertainty. 
The resulting values for the top quark mass are listed in Table~\ref{mtMSRfitTable}.

\begin{table}[t!]
\small
\centering
\caption{The values of $\mtMSR(R)$ obtained at different scales $R$ (given in
  brackets below $\mtMSR$), and the corresponding $\mtMSbar(\mtMSbar)$, the
  $\chi^2$ divided by the number of degrees of freedom $\Ndof$ in the fit,
  along with the fit and scale uncertainties for the $\mtMSR(R)$ extracted at
  $R=80\GeV$. The results are shown for the constant $\mur,\muf$ setting,
  where the central $\mur$ and $\muf$ values are set to $\mtMSR(80\GeV)$ in
  the whole $m_\ttbar$ distribution, and for the semi-dynamical (SD) setting
  where they are set to $\mtMSR(80\GeV)/2$ for $m_\ttbar < 420 \GeV$ and to
  $\mtMSR(80\GeV)$ for higher $m_\ttbar$. The fit and $\mur,\muf$
  uncertainties correspond to the MSR mass extracted at $R=80\GeV$. Within the
  reported accuracy, the uncertainty in the initial choice of $R$ agrees in
  all cases when the extracted $\mtMSR(R)$ is evolved to the reference $R$.
}
\renewcommand{\arraystretch}{1.2}
\label{mtMSRfitTable}
\begin{tabular}{@{}c c c c c c c c c @{}c@{}}
  ~
& ~
& \mtMSR 
& \mtMSR
& \mtMSR
& \mtMSbar 
& Fit 
& $\mur,\muf$  
& $R$\\
  $\mur,\muf$
& $\chi^2/\Ndof$
& {\footnotesize($80\GeV$)}
& {\footnotesize($1\GeV$)} 
& {\footnotesize($3\GeV$)} 
& (\mtMSbar)
& unc. 
& unc. 
& unc.\\
  setting
& ~
& $[\GeV]$
& $[\GeV]$ 
& $[\GeV]$ 
& $[\GeV]$
& $[\GeV]$
& $[\GeV]$
& $[\GeV]$
\\
\hline
  Const.
& 1.86/3
& $167.7$
& $173.2$
& $172.9$
& $163.3$
& $^{+0.6}_{-0.6}$
& $^{+0.4}_{-0.6}$
& $^{+0.4}_{-0.5}$\\
  SD
& 3.03/3
& $169.3$
& $174.8$
& $174.5$
& $164.8$
& $^{+0.5}_{-0.5}$
& $^{+0.2}_{-0.4}$
& $^{+0.2}_{-0.3}$
\end{tabular}
\end{table}
\renewcommand{\arraystretch}{1.0}
In particular, setting the central $\mur$ and $\muf$ to
$\mtMSR(80\GeV)$ and considering the complete $m_\ttbar$ distribution yields 
\begin{equation}
\mtMSR(1\GeV) = 173.2 \pm 0.6\,\text{(fit)}
                       ^{+0.4}_{-0.6}\,\text{($\mur,\muf$)}
                       ^{+0.4}_{-0.5}\,\text{($R$)} \GeV \,.
\label{mt_R80_1_1_1_1}
\end{equation}
The value for $\mtMSR(80\GeV)$ in this fit translates into $\mtMSbar(\mtMSbar) = 163.3^{+0.8}_{-1.0} \GeV$. 
This is compatible within uncertainties with the value of $\mtMSbar(\mtMSbar) = 162.1^{+1.0}_{-1.0} \GeV$ obtained at NLO in the ABMP16 5-flavor PDF set~\cite{Alekhin:2018pai}.

In accordance with the results shown in Fig.~\ref{xsec_300-333_GeV},
multiplying the scales $\mur$ and $\muf$ by 1/2 within $m_\ttbar < 420 \GeV$
is observed to increase the NLO cross section at $R=80\GeV$. To compensate for
this effect, the fit for $\mtMSR(80\GeV)$ leads to a somewhat larger value 
for the top quark MSR mass, reducing the predicted cross section especially in the
vicinity of the $\ttbar$ production threshold.  
This results in the value
\begin{equation}
\mtMSR(1\GeV) = 174.8 \pm 0.5\,\text{(fit)}
                        ^{+0.2}_{-0.4}\,\text{($\mur,\muf$)}
                        ^{+0.2}_{-0.3}\,\text{($R$)} \GeV.
\label{mt_R80_05_1_1_1}
\end{equation}
It is expected that the impact of the choices for $\mur$ and $\muf$, 
i.e. the shift of $1.6\GeV$ in the central values between Eqs.~\eqref{mt_R80_1_1_1_1} and \eqref{mt_R80_05_1_1_1}, 
will be reduced at NNLO accuracy and once a reliable description of the quasi-bound state effects is available. 
Nonetheless, as anticipated from the observations in Sec.~\ref{sec:scale_investigations}, the scale setting in Eq.~\eqref{mt_R80_05_1_1_1} already increases the robustness against scale variations, yielding somewhat smaller uncertainties than Eq.~\eqref{mt_R80_1_1_1_1}. 
Even though one may consider the $1.6\GeV$ difference as a quantification of the uncertainty of our current NLO analysis, a proper uncertainty analysis should include the NNLO and quasi-bound state corrections as well as order dependent input quantities (including PDFs, the strong coupling and their correlation) and is postponed to future work.

\begin{figure}[h!]
\centering
\includegraphics[width=0.495\textwidth]{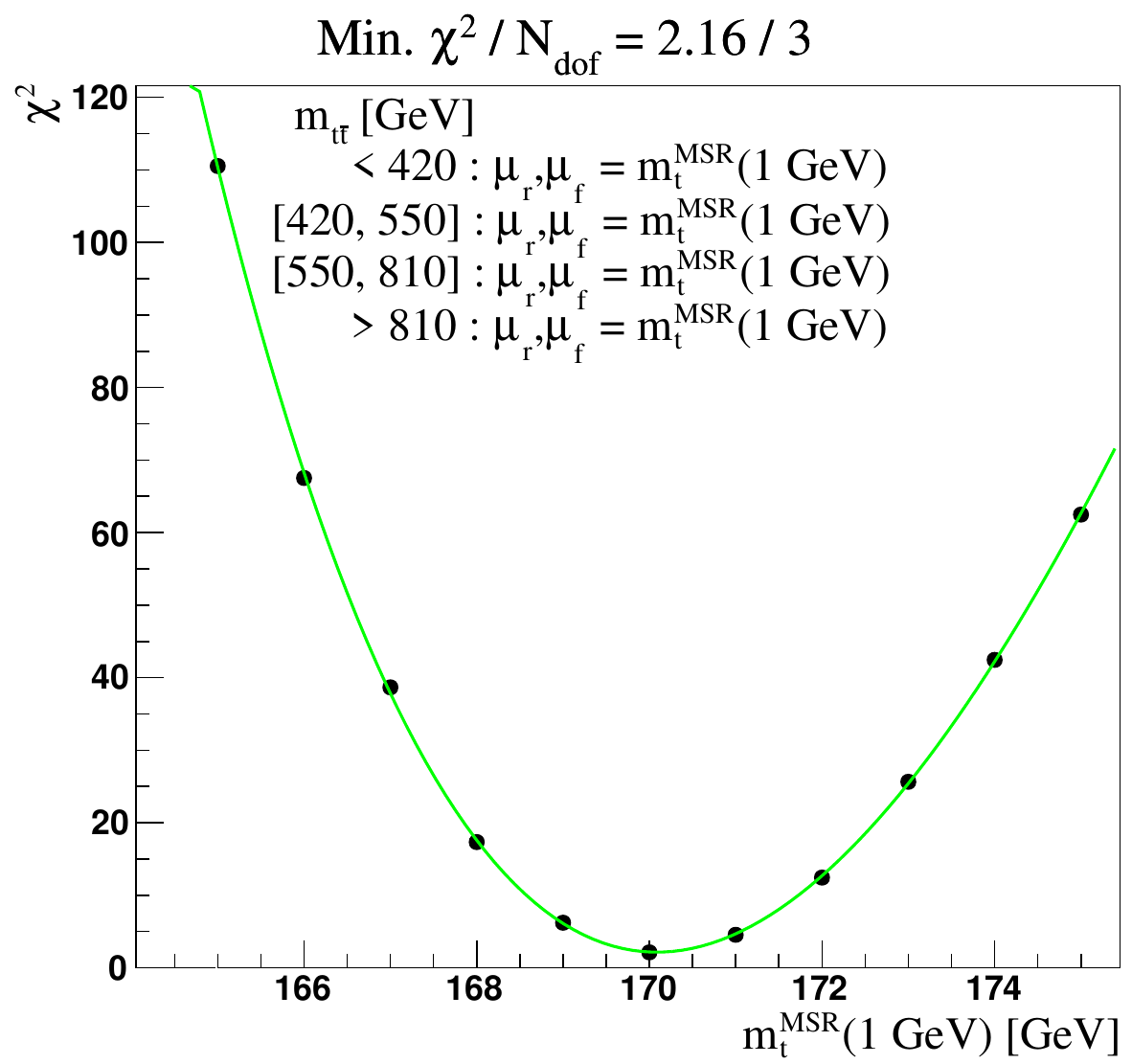}
\caption{Same as Fig.~\ref{mtFit_chi2}, now fitting $\mtMSR(1\GeV)$ and with the scales $\mur$ and $\muf$ set to $\mtMSR(1\GeV)$ in the whole $m_\ttbar$ distribution.}
\label{mtFit_R1_chi2}
\end{figure}
In order to illustrate the main conceptual novelty and the phenomenological importance of the mass scheme choice, we also perform the following variant of the fit:
Instead of determining the top quark MSR mass at $R=80\GeV$ and evolving the extracted $\mtMSR(80\GeV)$ value to $R=1\GeV$, as in Eqs.~\eqref{mt_R80_1_1_1_1} and \eqref{mt_R80_05_1_1_1}, 
we perform the fit to data directly with the initial scale set to $R=1\GeV$ in NLO cross section of Eq.~\eqref{diff_NLO_cs}.
Using also the central scales $\mur$, $\muf$ set to $\mtMSR(1\GeV)$, 
this results in 
\begin{equation}
\mtMSR(1\GeV) = 170.1 \pm 0.6\,\text{(fit)} ^{+1.1}_{-0.9}\,\text{($\mur,\muf$)} \GeV
\, ,
\label{mt_R1_1_1_1_1}
\end{equation}
where the corresponding fit to $\chi^2$ is shown in Fig.~\ref{mtFit_R1_chi2}. 
In Eq.~\eqref{mt_R1_1_1_1_1} the $\mur$ and $\muf$ scale uncertainties are twice as large as those of Eq.~\eqref{mt_R80_1_1_1_1}.
The even more sizeable discrepancy to the results of Eqs.~\eqref{mt_R80_1_1_1_1} and \eqref{mt_R80_05_1_1_1} indicates that scale variation does not provide a proper estimate of the theoretical uncertainties due to the missing higher order and quasi-bound state corrections for the result quoted in Eq.~\eqref{mt_R1_1_1_1_1}.
Since using $\mtMSR(1\GeV)$ closely approximates the outcome using pole mass scheme, this confirms our conclusions drawn in Sec.~\ref{sec:scale_investigations} that the use of the pole mass scheme (or a very small initial $R$ value for the MSR mass) leads to less reliable results in a fixed order QCD description at NLO accuracy, where the resummation of the quasi-bound state effects is missing. 
The significant difference of $4.7\GeV$ between the central values in  Eqs.~\eqref{mt_R80_05_1_1_1} and~\eqref{mt_R1_1_1_1_1}, as compared to the quoted scale uncertainties, demonstrates the phenomenological relevance of this issue.
This underpins the importance of an adequate top mass scheme choice as well as a proper scale setting in future $\mt$ extractions based on the $m_{\ttbar}$ measurements.

Let us now comment on other recent extractions of the top-quark mass, which have employed different methodologies.
Data from the CMS Collaboration for the $\ttbar$ production cross section
collected in $\pp$ collisions at the LHC at $\sqrt{s}=13\TeV$ 
have been used previously for a determination of the top-quark mass using both, 
the pole and the $\MSbar$ mass scheme~\cite{CMS:2018fks,CMS:2019esx}.
The emphasis of those analyses has been on keeping the correlations of the top-quark mass  with the strong coupling $\alpSZ$ and the PDFs.
In a different thread of analyses, the running of top quark $\MSbar$ mass 
$\mtMSbar(\mu_m)$ has been studied at NLO~\cite{CMS:2019jul} and 
NNLO~\cite{Defranchis:2022nqb} with dynamical scales, 
using data from the CMS Collaboration for the $m_\ttbar$ distributions.\footnote{See also
\url{http://cms-results.web.cern.ch/cms-results/public-results/publications/TOP-19-007/index.html\#Figure-aux_001}.}

Of these analyses, the results of Ref.~\cite{CMS:2019esx} can be compared to the present work, 
since they are obtained from normalized multi-differential cross sections which also include the low $m_\ttbar$ region discussed here, 
and the theoretical predictions are also based on the NLO \MCFM cross section description.
Ref.~\cite{CMS:2019esx} quotes $\mtpole = 170.5 \pm 0.8 \GeV$, which, if interpreted as the asymptotic pole mass~\cite{Hoang:2021fhn}, translates into $\mtMSR(1 \GeV) = 170.2 \pm 0.8 \GeV$.
This is compatible with the variant of the present study in Eq.~\eqref{mt_R1_1_1_1_1} obtained by directly fitting $\mtMSR(1 \GeV)$ to data, although the combined fit of 
 $\mtpole$, $\alpSZ$ and PDFs in Ref.~\cite{CMS:2019esx} reports a smaller value of $\alpSZ$ than used in Eq.~\eqref{mt_R1_1_1_1_1} on the basis of the ABMP16 PDF set, and a somewhat different gluon PDF.
The result of Ref.~\cite{CMS:2019esx} is systematically lower than those obtained using inclusive $\ttbar$ cross section measurements ~\cite{ATLAS:2015pfy,ATLAS:2019guf,CMS:2022emx}. The results of our analyses shown in Eqs.~\eqref{mt_R80_1_1_1_1} and \eqref{mt_R80_05_1_1_1} demonstrate that a larger and more reliable value could be obtained once the quasi-bound state, and potentially also NNLO corrections, are accounted for.

The value for the top quark MSR mass of $\mtMSR(3\GeV) = 169.6 ^{+0.8}_{-1.1}\GeV$, which corresponds to $\mtMSR(1\GeV) = 169.9 ^{+0.8}_{-1.1}\GeV$,  has been extracted in Ref.~\cite{Garzelli:2020fmd}, using the CMS data of Ref.~\cite{CMS:2019esx} and the same methodology, i.e. using fixed-order QCD perturbation theory at NLO accuracy,
so that $\mtMSR(3\GeV)$ has been fitted simultaneously with the PDFs and strong coupling constant.
Evolving the result of the present study in Eq.~\eqref{mt_R80_05_1_1_1} to $R=3\GeV$ yields
\begin{equation}
\label{mt_R80_05_1_1_1-evolved}
\mtMSR(3\GeV)=174.5 \pm 0.5\,\text{(fit)}
                    ^{+0.2}_{-0.4}\,\text{($\mur,\muf$)}
                    ^{+0.2}_{-0.3}\,\text{($R$)}\GeV \, ,
\end{equation}
which indicates some tension as well.\footnote{The computations in Ref.~\cite{Garzelli:2020fmd} rely on the
practical MSR (pMSR) definition~\cite{Hoang:2017suc} instead of the natural MSR (nMSR)
scheme used in this work. The difference is at the level of $10$~MeV~\cite{Hoang:2017suc} and thus negligible for the uncertainties quoted.} 
The result of Ref.~\cite{Garzelli:2020fmd} is very close to the one obtained in Ref.~\cite{CMS:2019esx} because using the MSR mass with a very low $R$ results in a value very close to the pole mass, as discussed above. The result obtained for the MSR mass $\mtMSR(3\GeV)$ in Ref.~\cite{Garzelli:2020fmd} and the differences obtained in our analyses shown in Eq.~\eqref{mt_R1_1_1_1_1} in comparison to Eqs.~\eqref{mt_R80_1_1_1_1} and \eqref{mt_R80_05_1_1_1}
underline that just using the MSR mass $\mtMSR(R)$ instead of the pole mass may not lead to a different result at all if the scale $R$ is not chosen in an adequate way. 
Further,
Ref.~\cite{Garzelli:2020fmd} has obtained 
$\alpSZ = 0.1132^{+0.0023}_{-0.0018}$, 
which is two standard deviations away from the value of the ABMP16 fit at NLO~\cite{Alekhin:2018pai} used in the extraction of Eq.~\eqref{mt_R80_05_1_1_1-evolved}. 
The similar values of $\mtMSR(3\GeV)$ in Ref.~\cite{Garzelli:2020fmd} and in Eq.~\eqref{mt_R1_1_1_1_1} indicate a less significant impact of the value of $\alpSZ$ on the $\mt$ extraction for the observable under consideration. A related observation was made for the top quark mass sensitivity in the 2-jettiness distribution analyzed in Refs.~\cite{Butenschoen:2016lpz} for boosted top pair production $e^+e^-$ collisions.

Notably, neither any of the cited previous top quark mass extractions based on the LHC data
nor the present work have included the aforementioned corrections for the quasi-bound state effects. 
However, the extraction of the top quark MSR mass using predictions in the MSR scheme at the scale $R=80\GeV$ profits from the smaller size of these effects and thus from an improved stability of the cross section.

We note that the ATLAS Collaboration has derived a value for the top quark MSR mass at the reference scale $R=1\GeV$ in Ref.~\cite{ATLAS:2021urs} 
by comparing QCD predictions at next-to-leading logarithmic accuracy for the 
soft-drop groomed top quark jet mass distribution 
to parton shower Monte Carlo simulations for a Monte-Carlo top quark mass $\mt^{\rm MC}=172.5\GeV$.
Obtained in the Monte Carlo calibration (following~\cite{Butenschoen:2016lpz}), the result of Ref.~\cite{ATLAS:2021urs} is not based on experimental data and hence cannot be directly compared to the results of the present study.

\section{Summary and Conclusions}
\label{sec:summary}

We have presented the first comprehensive study of the $m_\ttbar$ distribution in its dependence on
the mass renormalization scales $R$ and $\mu_m$ of the MSR and $\MSbar$ top quark mass schemes. 
Our findings suggest that using the MSR mass $\mtMSR(R)$ with the scale setting of $R$ close to $80\GeV$ improves 
the robustness of the predictions for the $m_\ttbar$ distribution 
against scale variations in general and, in particular, against the impact of quasi-bound state corrections 
in the region of $m_\ttbar$ close to the $\ttbar$ threshold. 
The theory predictions used in our study are based on the NLO fixed order QCD description provided by the \MCFM program, adapted to the MSR and $\MSbar$ top quark mass schemes.
The optimized scale choices for those mass schemes are characterized by low values of the renormalization and factorization scales $\mur$ and $\muf$.
This holds in particular in the vicinity of the $\ttbar$ production threshold region in the $m_\ttbar$ distribution, where values $\mur \simeq \muf \simeq \mt/2$ are observed to stabilize cross section predictions and to decrease the scale uncertainty in the determination of the MSR mass. 

These settings have been applied in an extraction of the 
top quark MSR mass at $R=80\GeV$, using $\ttbar$
pair production cross section, measured as a function of $m_{\ttbar}$ in
$\pp$ collisions at $\sqrt{s}=13\TeV$ at the LHC by the CMS Collaboration, 
using fixed-order perturbative QCD predictions at NLO accuracy and also the semi-dynamical scales for $\mur,~\muf$ in the low-$m_\ttbar$ regime.
The fitted value of $\mtMSR(80\GeV)$ has then been evolved to various low reference scales $R$, rather than computing the cross sections directly at low $R$ as performed in earlier analyses.
This procedure yields the value $\mtMSR(3\GeV)=174.5^{+0.6}_{-0.7}\GeV$, which is discussed in the context of other recent extractions of the top quark mass from LHC data. 
The observed differences are explained to a large part by the scale choice of $R=80\GeV$ for the top quark MSR  mass, advocated by the present study, signifying the achieved systematic resummation of higher order QCD corrections.
Other reasons for differences are due to the choice of the value for the strong coupling $\alpSZ$, which directly affects the normalization of the cross section and is anti-correlated with the top quark mass, and, to a lesser extent, due to the particular PDF sets used.

While we have argued that the implementation of the MSR mass scheme in the $\ttbar$ cross section calculation and the optimal scale choice for $R$ of $80\GeV$ provide more robust
predictions even at NLO accuracy, the findings should be corroborated by extending the analysis to NNLO accuracy.
In addition, the proper treatment of both, the quasi-bound state effects, together with a matching to the relativistic $\ttbar$ region, and the $m_\ttbar$ region below the threshold are further important improvements to be implemented. 
A final reliable measurement of the top quark MSR mass needs to address those issues as well as the correlation of the top quark mass with the other theoretical parameters, which control the cross section predictions. 
We leave these aspects for future studies.


\section*{Acknowledgements}

The work of A.H.H was supported in part by FWF Austrian Science Fund under the Project No.\ P32383-N27,
the work of S.M. in part by the Bundesministerium f{\"u}r Bildung und Forschung under contract 05H21GUCCA, the work by T.M. and K. L. is supported by the Helmholtz Association under the contract W2/W3-123, and T.M. is also supported by the National Science Centre, Poland, research grant No. 2021/42/E/ST2/00031.


\printbibliography[heading=bibliography]

\end{document}